\documentclass[fleqn,usenatbib]{mnras}

\usepackage[T1]{fontenc}
\usepackage{ae,aecompl}

\usepackage{graphicx}	
\usepackage{amsmath}	
\usepackage{amssymb}	

\title[Type Ia SN from wide TWDs] {Type Ia Supernovae from wide white-dwarfs triples}

\author[E. Michaely]{
Erez Michaely$^{1}$\thanks{E-mail: erezmichaely@gmail.com}
\\
$^{1}$Astronomy Department, University of Maryland, College Park, MD 20742\
}
\date{Accepted XXX. Received YYY; in original form ZZZ}

\pubyear{2020}

\begin{document}
\label{firstpage}
\pagerange{\pageref{firstpage}--\pageref{lastpage}}
\maketitle

\begin{abstract}

For ultra-wide systems (with outer orbit >$10^{3}{\rm AU})$ the galactic
field is collisional. Hence, ultra-wide triple white-dwarfs (TWDs)
can be perturbed, by flyby stars, to sufficiently high outer eccentricity
such that the triple becomes dynamically unstable. An unstable triple
undergoes multiple binary-single resonant encounters between all three
WDs. These encounters might result in a direct collision between any
random two WDs and lead to a Type Ia supernova (SN) event. In case
where the multiple resonant encounters did not produce a collision
a compact binary is formed (while the third WD is ejected), this binary
either collides or merges via gravitational wave emission, similar
to the classic double-degenerate (DD) channel. In this research study we estimate
the galactic rates of Type Ia SN from the direct collision channel is to be $0.1\%-4\%$
and primarily  $2\%-36\%$ from the DD scenario.
\end{abstract}

\section{Introduction}
\label{sec:Introduction}
One of the great open questions in astrophysics today is regarding
the origin of Type Ia supernovae (SNe) \citep{Maoz2014,Livio2018,Soker2019,Ruiter2020}. A type Ia
SN is believed to be a thermonuclear explosion of a white-dwarf (WD)
\citep{Hoyle1960}, however the channel or channels that lead to the
explosion are under active research and debated frequently. SNe Ia
are classified as such by a lack of hydrogen and helium lines in their
spectra and by strong and wide lines of silicon, iron and calcium.
Unlike core-collapse SNe \citep{Smartt2009} which originate from
massive (young) stars, thus only observed in star forming galaxies,
Type Ia SNe are observed both in young (spiral galaxies) and old (elliptical
galaxies) environments \citep{Maoz2014}.

Not all Ia SNe events are similar, there is a subclass of events named
``peculiar SNe'' which exhibits different characteristics. Their
peak luminosity is usually lower than the standard events, like the 1991bg-like SNe; however,
there were over-luminous cases observed, e.g. SN 1991T-like \citep{Ruiter2020}. Peculiar
SNe are hosted primarily in star forming galaxies and evolve faster
than their ``regular SNe'' counterparts. In this manuscript we focus
on regular Type Ia which are powered by the explosion of a carbon-oxygen
(CO) WD.

There are several classical channels that aim to explain the evolution
that lead to a Type Ia SN, we list them with no specific order.

The two popular theories are the single-degenerate (SD) and the double-degenerate
(DD) theories. In the SD theory \citep{Whelan1973,Nomoto1982} a CO WD accretes mass from a non-degenerate binary companion, sufficiently
and effectively to reach the Chandrasekhar mass limit, and consequently
explodes. In the DD theory \citep{Iben1984,Webbink1984} two WDs coalesce
through gravitational wave (GW) emission. The merged product reaches the Chandrasekhar mass
limit and explodes.

A third channel is the ``core-degenerate'' (CD) theory, where a
degenerate WD merges with the hot core of an AGB-star, exceeds the Chandrasekhar
mass limit and explodes when the core sufficiently cools \citep{Kashi2011,Ilkov2012}.
A fourth channel is the ``double-detonation'' (DDet) \citep{Woosley1994,Livne1995,Shen2018},
in this sub-Chandrasekhar channel a WD accretes a layer of helium-rich
material from a binary companion. The helium layer is ignited and
detonates when compressed under the accreted material leading to a
second detonation near the center of the CO WD. A recent scenario,
fifth, describes the tidal disruption of a hybrid WD by a CO WD \citep{Perets2019}.
Hybrid WDs are a natural outcome of binary stellar evolution and often
overlooked as a possible source for Type Ia SNe. Similar to the DD
theory, a hybrid WD can be tidally disrupted by the gravitational
field of a close CO WD and form an accretion disk that detonates.
All of the theories above present strong and week points \citep{Maoz2014,Tsebrenko2015,Soker2019}
such that none of these theories are in consensus of the community.

The sixth and final theory we review here is the WD-WD collision
scenario (2WDC) \citep{Raskin2009,Thompson2011,Katz2012,Kushnir2013}.
In this scenario two WDs undergo direction collision, usually with
a presence of a third companion and immediately ignite. The main strong
point of this scenario is the well understood exploding mechanism
\citep{Kushnir2013} which is lacking from the other theories. However,
this channel suffers from an extremely low merger rates \citep{Toonen2018,Hamers2013,Prodan2013}
and can explain <0.1\% of observed SN rate.

In this paper we revisit this scenario and describe a new channel
that may explain up to $~4\%$ of the observed rate of
Type Ia SNe from direct collisions and addition up to $~36\%$ from the new channel to the DD scenario .

Recently it was shown that for wide (semi-major axis $>1000{\rm AU}$)
systems (either for binaries or triples) the field of the galaxy is collisional
\citep{Kaib2014,Michaely2016,Michaely2019,Michaely2020}. During the
lifetime of wide systems multiple gravitational interactions occur
between the systems and a stellar flybys. This results in a change
in the outer eccentricity of the outer binary, in the case of a triple
or the eccentricity of the wide binary itself. \citet{Michaely2016}
showed that this may explain the formation of low-mass X-ray binaries,
or gravitational waves sources from wide binary-BH in the field \citep{Michaely2019}
or recently from triple BH systems \citep{Michaely2020}.

In this manuscript we mainly follow the dynamical treatment in \citet{Michaely2020}
and calculate the collision rate of two WD originating from initially
wide triple systems.

This paper is organized as follows: In section \ref{sec:Wide-triples-in}
we describe the interaction of the triple in the field and calculate
the fraction of systems that undergo the instability phase. In section
\ref{sec:Unstable-triples} we treat the instability phase were multiple
binary-single encounters occur, including the endstate. In section
\ref{sec:SNe-rates} we calculate the Type Ia SN rate for spiral and
elliptical galaxies. Section \ref{sec:Discussion} is dedicated to
discussing the results and future work while we conclude the research
in the summary in section \ref{sec:Summary}.

\section{Field Wide WD triples}
\label{sec:Wide-triples-in}
In the following we describe the imuplse interaction between a random flyby star and
a wide triple WD system in the field of the host galaxy. Similar mathematical description
can be found in previous papers \citep{Michaely2016,Michaely2019,Michaely2020}.
However, here we emphasize the main differences of this study from \citep{Michaely2020}.
Namely, the main goal in \citep{Michaely2020} is to calculate the
merger rate of binary BH via gravitational wave emission, while here
we focus on direct collision between two WDs. However, in section
\ref{subsec:Mergers-in-theES} we estimate the merger of two WDs
via GW emission, similar to the standard DD scenario. We first describe
the interaction in subsection \ref{subsec:Qualitative-description}
and treat the interaction ina quantitative way in section \ref{subsec:Quantitative-description}.

\subsection{Qualitative description}
\label{subsec:Qualitative-description}
\begin{figure}
\includegraphics[width=1\columnwidth]{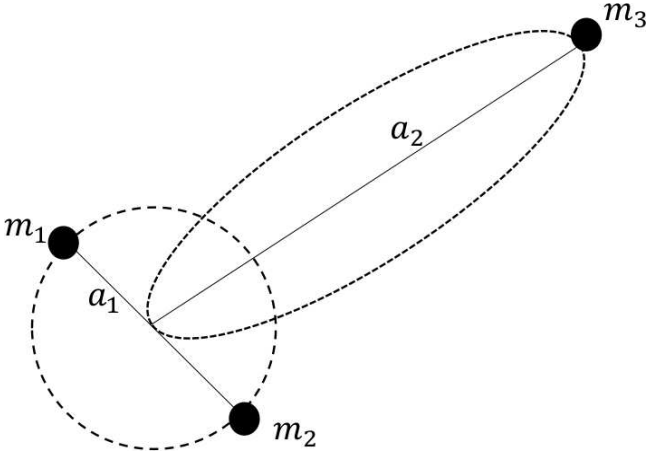}\caption{\label{fig:Illustration-of-hierarchical}Cartoon of wide hierarchical
TWD, $a_{1}\ll a_{2}$. The inner binary orbit is circular. The outer binary eccentricity is drawn from a thermal distribution,
$f\left(e_{2}\right)=2e_{2}.$}

\end{figure}

It was previously shown by\citep{Kaib2014,Michaely2016,Michaely2019,Michaely2020}
that wide systems, either binaries of triples with outer semi-major
axis (SMA), $a\gtrsim1000{\rm AU}$, interact with flyby stars sufficiently
to change their pericenter distances, mainly through change in their
eccentricity \citep{Lightman1977,Merritt2013}. This might lead to
either a tidal interaction \citet{Michaely2016}, or inspiral due to
GW emission \citep{Michaely2019} or destabilize the triple \citep{Michaely2020}.
In this manuscript we focus on triple WDs
(TWDs) in a wide and hierarchical configuration, where all the WD masses are
equal; the inner binary is constructed from two masses, $m_{1}$ and $m_{2}$.
The inner SMA is denoted by $a_{1}$ and the inner binary eccentricity 
is $e_{1}$ and is set to zero, for simplicity. The third WD, $m_{3}$
and the center of mass of the inner binary is considered as an outer
binary with SMA, $a_{2}$ provided that $a_{2}\gg a_{1}$. For illustration
see Figure \ref{fig:Illustration-of-hierarchical}. One can relax
the assumption that all three objects are WDs and estimate the collision
rate of any triple that consists two WDs. However, this would be dominated
by the collision of a WD with the third stellar companion, due to
the size of the stellar companion with respect to a WD. This will
be the focus of a future research.

Similar to \citep{Michaely2020} we are neglecting Lidov-Kozai oscilations effects
\citep{Lidov1962,koz62,Naoz2016,Michaely2014}. Here we focus
on wide systems with $a_{2}>1000{\rm AU}$. Therefore the
Lidov-Kozai timescale is 
\begin{equation}
\tau_{{\rm LK}}\approx\frac{P_{2}^{2}}{P_{1}}\approx
\end{equation}
 
\[
6.6\cdot10^{13}{\rm yr}\left(\frac{a_{2}}{10^{4}{\rm AU}}\right)^{3}\left(\frac{a_{1}}{0.1{\rm AU}}\right)^{-\frac{3}{2}}\left(\frac{M}{1.8M_{\odot}}\right)^{-1}\left(\frac{M_{b}}{1.2M_{\odot}}\right)^{\frac{1}{2}}
\]
where $M\equiv m_{1}+m_{2}+m_{3}$ is the total mass of the triple
and $M_{b}\equiv m_{1}+m_{2}$ is the total mass of the inner binary,
$\tau_{{\rm LK}}$ is larger than Hubble time.

A flyby interaction with wide systems can change the outer binary eccentricity 
in a way that the pericenter passage is inside the SMA of the inner binary, $q=a_{2}\left(1-e_{2}\right)\lesssim a_{1}$.
In this case, the triple becomes dynamically unstable \citep{Stone2019b,Samsing2014,Heggie1975}
and as a consequence the triple breaks down to a series of binary-single encounters, similar to the dynamics expected in dense
environments even-though the triple resides in the field of the galaxy.
In other words the dynamics of an unstable triple in the field is
similar to the dynamics in dense environments.

The are two important timescales for flyby-triple interaction. First,
the flyby-triple interaction timescale, $t_{{\rm int}}\equiv b/v_{{\rm enc}}$, 
where $b$ is the closest approach of the stellar flyby to the center of mass of the triple system, and $v_{{\rm enc}}$ is
the relative velocity at infinity of the stellar flyby compared to the center of mass of the triple. 
Second timescale is the orbital period of the outer binary, $P_{2}$. Here, we consider only the impulsive interaction regime where $t_{{\rm int}}\ll P_{2}$.
In section \ref{subsec:Quantitative-description} we calculate the
rate where unstable triples are formed from hierarchical triples as a function
of the inner and outer SMA.

During the dynamical instability phase multiple binary-single encounter
occur. In every encounter a temporary binary is created with SMA,
$a'$ and eccentricity, $e'$. This binary is bounded to the third
WD which orbits the center of mass of the temporary binary in a Keplerian
orbit with timescale of $t_{{\rm iso}}$ \citep{Michaely2020}. A
fraction of all systems experience a direct collision between the
two components of the inner temporary binary, $f_{{\rm collision}}\left(a_{1}\right)$.
We find $f_{{\rm collision}}\left(a_{1}\right)$ in section \ref{subsec:Calculating-the-merger}.

\subsection{Quantitative description}
\label{subsec:Quantitative-description}
Consider an ensemble of hierarchical and wide TWDs. The masses of the WDs are equal $m_{1}=m_{2}=m_{3}=0.6M_{\odot}$
(we denoted the total mass of the system by $M$) the inner (outer)  SMA $a_{1} \left(a_{2}\right)$.
The distributions of $a_{1}$ and $a_{2}$ are log-uniform,
$f_{a_{1(2)}}\propto1/a_{1(2)}$ between $0.01{\rm AU-100{\rm AU}}\!\left(10^{3}{\rm AU}-10^{5}{\rm AU}\right)$.
The inner binary is set to be circular, $e_{1}=0$ whereas the outer binary eccentricity
is drawn from a thermal distribution, $f\left(e\right)de=2ede.$
All triples in the ensemble are embedded in the field with number stellar density of
$n_{*}$ and a  velocity dispersion of $\sigma_{v}$ which we
set to be the relative velocity at infinity, $v_{{\rm enc}}$.

Next we derive the fraction of the ensemble that, $q\leq a_{1}$. Namely, 
the third WD inside the SMA of the inner binary, and write it as a function of 
the outer SMA, $a_{2}$ and field number
density, $n_{*}.$ Moreover, we account for outer binary ionization
from the random interaction with flyby stars.

The loss cone, $F_{q}$, is the fraction out of the systems in the ensemble
that the $q\leq a_{1}$. The equality $q=a_{1}$ defines the critical
eccentricity $e_{c}$ where the system is marginally stable, namely 
\begin{equation}
a_{2}\cdot\left(1-e_{c}\right)=a_{1}
\end{equation}
 which corresponds to $e_{c}=1-a_{1}/a_{2}$. 
\begin{equation}
F_{q}=\int_{e_{c}}^{1}2ede=\frac{2a_{1}}{a_{2}}.\label{eq:Fq}
\end{equation}
The loss cone is small, namely $F_{q}\ll1$. If a TWD is in the loss cone, $m_{3}$
enters the inner binary within the outer orbital period, $P_{2}$
and a dynamical instability begins. Then the system is lost from the
ensemble. Systems with eccentricities close to the critical eccentricities could potentially
be perturbed to enter the loss cone and replenish it after the next flyby
happens. In order to calculate what is the fraction of systems that are susceptible to enter the loss cone we calculate the smear cone. 
The smear cone is a measure in phase space that an outer binary can occupy after an impulsive interaction
with a random flyby star. Defined by $\theta=\left\langle \Delta v\right\rangle /v_{k}$,
 where $v_{k}$ is the relative Keplerian velocity between the components of the outer binary at
the average separation, $\left\langle r\right\rangle =a_{2}\left(1+1/2e^{2}\right)$.
Because $F_{q}\ll1$ we approximate $e\rightarrow1$, namely $v_{k}=\left(GM/3a_{2}\right)^{1/2}$,
where $G$ is Newton's constant. The change in velocity $\Delta v\approx3Ga_{2}m_{p}/v_{{\rm env}}b^{2}$
\citep{Hills1981,Michaely2019} where $m_{p}$ denotes the mass of the
stellar flyby. Following \citep{Michaely2019} we write can the
smear cone 
\begin{equation}
F_{s}=\frac{\pi\theta^{2}}{4\pi}=\frac{27}{4}\left(\frac{m_{p}}{M}\right)^{2}\left(\frac{GM}{a_{2}v_{{\rm enc}}^{2}}\right)\left(\frac{a_{2}}{b}\right)^{4}.\label{eq:SmearCone}
\end{equation}
The fraction of the loss cone filled after a single flyby interaction is given by the ratio of the smear cone to loss cone:
\begin{equation}
\frac{F_{s}}{F_{q}}=\frac{27}{8}\left(\frac{m_{p}}{M}\right)^{2}\left(\frac{GM}{a_{2}v_{{\rm enc}}^{2}}\right)\left(\frac{a_{2}}{b}\right)^{4}\left(\frac{a_{2}}{a_{1}}\right).
\end{equation}
For the situation where the loss cone is continuously fill, $F_{q}=F_{s}$,
the timescale of which systems are depleted is just the outer binary
orbital period, $P_{2}$. Therefore one can write the depletion rate, which is
a function of the size of the loss cone to be
\begin{equation}
\dot{L}_{{\rm Full}}=\frac{F_{q}}{P_{2}}\propto a_{2}^{-5/2}a_{1}
\end{equation}
notice that the depletion rate is independent of the local stellar density, $n_{*}$ (the position in the galaxy) and scales
 with the inner binary SMA, $a_{1}$. Hence the depletion
rate is decreasing with wider outer SMA in the full loss cone regime.

\begin{figure}
\includegraphics[width=1\columnwidth]{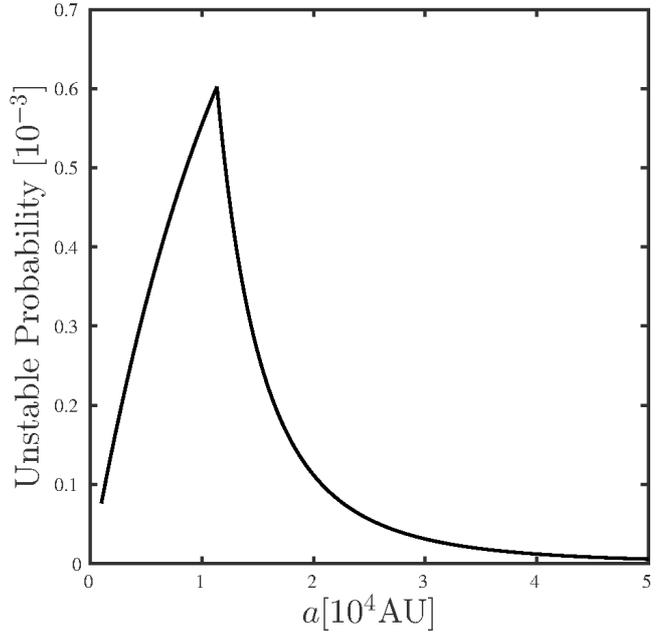}\caption{\label{fig:Probability-of-becoming}The probability for the triple
to become unstable due to flyby interaction. Namely, the outer pericenter
distance $q_{2}=a_{2}\left(1-e_{2}\right)\protect\leq a_{1}$. The
plot is calculated for the following parameters: $t=10{\rm Gyr}$,
$n_{*}=0.1{\rm pc^{-3}},$$v_{{\rm enc}}=50{\rm kms^{-1}}$. The highest
probability is for $a_{{\rm crit}}$. The full loss cone regime is
for $a>a_{{\rm crit}}$ and the empty loss cone regime is for $a<a_{{\rm crit}}.$}

\end{figure}

In the case, where $F_{s}<F_{q}$, the loss cone is not completely
full all the time. Namely for outer binaries with tighter orbits, which are less susceptible for eccentricity change from a random flyby
interaction (\ref{eq:SmearCone}), we term this situation as the empty loss cone regime. In this case
the depletion rate depends on the rate of systems being kicked into
the loss cone. Specifically, $f=n_{*}\sigma v_{{\rm enc}}$ where
$\sigma=\pi b^{2}$ is the geometric cross-section of the random flyby
interaction. In this case the timescale where the depletion occur
is the timescale for entering the loss cone, namely $T_{{\rm empty}}=1/f$.
We can find following this equation that $f$ is just:
\citep{Michaely2016,Michaely2019} 
\begin{equation}
f=n_{*}\pi\sqrt{\frac{27}{8}\left(\frac{m_{p}}{M}\right)^{2}\frac{GMa_{2}^{4}}{a_{1}}}.
\end{equation}
When the two timescales are equal the rate where systems leave the loss cone is equal to the
rate of systems entering the loss cone. The critical SMA that separates
the full lose cone regime to the empty lose cone regime (where the
timescales are equal \citep{Michaely2019}) is given by
\begin{equation}
a_{{\rm crit}}=\left(\frac{2}{27\pi^{4}}\frac{M}{m_{p}^{2}}\frac{a_{1}}{n_{*}^{2}}\right)^{1/7}.
\end{equation}
 Using $a_{{\rm crit}}$ we can calculate the fraction of systems that
enter the loss-cone for the empty loss cone regime: $a<a_{{\rm crit}}$, and for the full
loss cone, $a>a_{{\rm crit}}$.
$F_{q}$ is the loss fraction of systems in the ensemble after the relevant timescale, hence $\left(1-F_{q}\right)$
is the fraction of the surviving systems. The relevant timescale for the empty loss cone regime 
is $T_{{\rm empty}}=1/f$. For the full loss cone regime that timescale
is $P_{2}$, the outer orbital period. Next we write the fraction of systems that enter the loss
cone as a function of time, $t$ as 
\begin{equation}
L\left(a_{1},a_{2},n_{*}\right)_{{\rm empty}}=1-\left(1-F_{q}\left(a_{1},a_{2}\right)\right)^{t\cdot f}.
\end{equation}
For the limit $F_{q}t/T_{{\rm empty}}\ll1$ we can take the leading term to get the approximation 
\begin{equation}
L\left(a_{1},a_{2},n_{*}\right)_{{\rm empty}}=F_{q}tf
\end{equation}
which is proportional to the size of the loss-cone, $F_{q}$, specifically 
\begin{equation}
L_{{\rm empty}}\propto F_{q}\propto a_{2}^{-1}a_{1}.
\end{equation}
The loss fraction grows with SMA for $a_{2}<a_{{\rm crit}}$,
unlike the full loss cone regime. This means that the highest lost
fraction comes from TWD with SMA of $a_{{\rm crit}}.$ For the full
loss cone we make the same treatment with 
\begin{equation}
L\left(a_{1},a_{2},n_{*}\right)_{{\rm full}}=1-\left(1-F_{q}\left(a_{1},a_{2}\right)\right)^{t/P_{2}},
\end{equation}
and after expansion we get 
\begin{equation}
L\left(a_{1},a_{2},n_{*}\right)_{{\rm full}}=F_{q}t/P_{2}.
\end{equation}
In the above mathematical treatment we ignored the disruption process for wide systems
in collisional environments. Taking this into account by
calculating the half-life time of the wide system from \citep{Bahcall1985}, where
the half-life time is  
\begin{equation}
t_{1/2}=0.00233\frac{v_{{\rm enc}}}{Gm_{p}n_{*}a_{2}}\label{eq:ionization}
\end{equation}
accounting for the ionization process in the empty loss-cone we get
\begin{equation}
L\left(a_{1},a_{2},n_{*}\right)_{{\rm empty}}=\tau F_{q}f\left(1-e^{-t/\tau}\right)=\label{eq:empty}
\end{equation}
\[
\tau\frac{2a_{1}}{a_{2}}n_{*}\pi\sqrt{\frac{27}{8}\left(\frac{m_{p}}{M}\right)^{2}\frac{GMa_{2}^{4}}{a_{1}}}\left(1-e^{-t/\tau}\right).
\]
where $\tau=t_{1/2}/\ln2$. While for the full loss-cone case:
\begin{equation}
L\left(a_{1},a_{2},n_{*}\right)_{{\rm full}}=\tau\frac{F_{q}}{P_{2}}\left(1-e^{-t/\tau}\right)=\label{eq:full}
\end{equation}
\[
\tau\frac{2a_{1}}{a_{2}}\left(\frac{GM}{4\pi^{2}a_{2}^{3}}\right)^{1/2}\left(1-e^{-t/\tau}\right).
\]
As mentioned in previous work \citep{Michaely2020} the lost fraction is proportional
to $a_{1}$. A representative example of the loss probability, or the probability that a TWD becomes unstable due to flyby interactions is presented in
figure \ref{fig:Probability-of-becoming}. 

In the next section we calculate the fraction of the cases where the triple becomes unstable and ends up as a Type Ia SN. This can happen either in a direct collision or 
a merger via GW emission.

\section{Instability stage}
\label{sec:Unstable-triples}
When the triple enters the loss cone it becomes unstable. In the follow section we describe the dynamics this instability. Similar treatment is done in  \citep{Samsing2014,Samsing2018b,Michaely2020}.

The triple losses it hierarchy and as a consequence multiple binary-single encounters occur. 
The physics of binary-single encounters were studied
mainly in the context of dense stellar environments,e.g. globular-clusters or galactic
centers. A close binary-single interaction occurs when the
single component passes within the binary sphere
of influence. For a close binary-single interaction every gravitational force exerted 
between every pair of components is comparable in strength, hence the outcome
is chaotic. Only two possible outcomes for the close binary-single interaction. First, direct-interaction,  only one gravitational interaction
takes place and the result is a binary, on a compact orbit and an single escaper. Second, intermediate state (IMS) or a resonant
phase, where the triple turns to a temporary binary and a bound third object on a wide orbit. This happens multiple times ($\left\langle N_{{\rm IMS}}\right\rangle =20$),
\citep{Samsing2017} for every binary-single encounter
the orbital elements of both binaries (inner and outer) are drawn from
the available distributions that conserve angular momentum and
energy. The end-state of the multiple binary-single similar to the direct-interaction case, where a tight binary
if formed, from any random two objects, and the third object escapes to infinity, an escaper.

From a direct collision perspective an event can occur either between
scatters in the IMS or after the end-state is reached when the final
binary escapes the gravitational potential of the third mass. In this study we calculate the merger rate only in the end-state case because we neglect the tiny fraction where
 the inner binary system losses enough orbital energy during the IMS to disconnect itself from the triple. In the
following we calculate the rate of collisions in both cases and the merger rate from GW emission.

\subsection{Intermediate state}

Next we model the dynamics of the IMS,
while in section \ref{subsec:Post-resonance-state} we describe the
endstate in the post resonant phase. In this study we only consider
WDs with equal masses of $0.6M_{\odot}$. The initial binary has a 
SMA, $a_{1}$, and zero eccentricity. The single (third) WD interacts gravitationally with the inner binary
, a close binary-single encounter. For each encounter a temporary binary is formed (with two random components out of the three WDs).
The temporary binary eccentricity, $e_{{\rm IMS}}$ is drawn from thermal
distribution and the SMA, $a_{\rm IMS}$ is determined
by the energy budget which is approximated by equation 12 in \citep{Samsing2018b}
\begin{equation}
\frac{m_{1}m_{2}}{2a_{1}}=\frac{m_{i}m_{j}}{2a_{{\rm IMS}}}+\frac{m_{ij}m_{k}}{2a_{{\rm bs}}}\label{eq:energy budget}
\end{equation}
where $a_{{\rm bs}}$
is the temporary SMA of the outer binary. Where $\left\{ i,j,k\right\} $
are the randomize indexes after the interaction and $m_{ij}=m_{i}+m_{j}$
is the mass of the temporary binary. From eq. (\ref{eq:energy budget})
the outer SMA of the third bound WD can be written as
\begin{equation}
a_{{\rm bs}}=a_{1}\left(\frac{m_{ij}m_{k}}{m_{1}m_{2}}\right)\left(\frac{a'}{a'-1}\right)\label{eq:a_bs}
\end{equation}
where 
\begin{equation}
a'\equiv\frac{a_{{\rm IMS}}}{a_{c}}\ {\rm and}\ a_{c}\equiv a_{1}\frac{m_{i}m_{j}}{m_{1}m_{2}}.\label{eq:a_definitions}
\end{equation}
In the equal mass case $a_{c}=a_{1}$ and therefore $a'$
is just $a_{{\rm IMS}}/a_{1}$.

We write the lower and upper bound of $a'$. The lower bound of $a'$ is trivial with 
\begin{equation}
a_{{\rm L}}'\approx1,\label{eq:a_L}
\end{equation}
while the upper bound should separate between the cases in the resonant state when the triple well 
described as a binary and a bound single, or when no temporary binary could be defined. This
occurs, $a_{{\rm bs}}\approx a_{{\rm IMS}}$. \citet{Samsing2018b}
found that one way of estimating $a'_{{\rm U}}$ is by comparing the
tidal force, $F_{{\rm tid}}$ exerts by the third BH to the binary
gravitational biding force, $F_{{\rm bin}}$. In the high eccentricity
limit we find 
\begin{equation}
F_{{\rm tid}}\approx\frac{1}{2}\frac{Gm_{ij}m_{k}}{a_{{\rm bs}}^{2}}\frac{a_{{\rm IMS}}}{a_{{\rm bs}}}
\end{equation}
\begin{equation}
F_{{\rm bin}}\approx\frac{1}{4}\frac{Gm_{i}m_{j}}{a_{{\rm IMS}}^{2}}.
\end{equation}
We set $a'_{{\rm U}}$ by the case that 
\begin{equation}
\frac{F_{{\rm tid}}}{F_{{\rm bin}}}=0.5
\end{equation}
which translates to 
\begin{equation}
a'_{{\rm U}}=1+\left(\frac{1}{2}\frac{m_{k}}{\mu_{{\rm ij}}}\right)^{2/3}\label{eq:a_U}
\end{equation}
where $\mu_{ij}\equiv m_{i}m_{j}/(m_{i}+m_{j})$ is the reduced mass
of the IMS binary.

The temporary inner SMA,  $a'$ values are distributed uniformly between $a'_{{\rm L}}$
and $a'_{{\rm U}}$ and the eccentricity distribution is thermal \citep{Heggie1975,Hut1985,Rodriguez2018}.

Given the temporary SMA, $a'$, and eccentricity, $e'$, one can calculate
the pericenter $q'=a'\left(1-e'\right)$ and compare that to the combine
radii of the two WD via the mass radius relation given by \citep{Carroll2006}
\begin{equation}
R_{{\rm WD}}=2.9\times10^{8}\text{\ensuremath{\left(\frac{M_{{\rm WD}}}{M_{\odot}}\right)}}^{-1/3}\left[{\rm cm}\right],\label{eq:M-R_relation}
\end{equation}
where if $q'\leq2R_{{\rm WD}}$ we flag it as a direct collision.
In the case where a collision did not occur we can calculate the orbital
time of the third companion, $t_{{\rm iso}}$. The orbital period
is simply the Keplerian orbital period with $a_{{\rm bs}}$, combining
it with eq. (\ref{eq:a_bs}) and eq. (\ref{eq:a_definitions}) we
get:
\begin{equation}
t_{{\rm iso}}=2\pi\frac{a_{1}^{3/2}}{\sqrt{GM}}\left(\frac{m_{ij}m_{k}}{m_{1}m_{2}}\right)^{3/2}\left(\frac{a'}{a'-1}\right)^{3/2}.\label{eq:t_iso}
\end{equation}

In section \ref{subsec:Post-resonance-state} we treat the case where
no collision happen during the $\left\langle N_{{\rm IMS}}\right\rangle $
scatters, and the end result is a compact binary and an escaper third
object.

\subsubsection{Estimating the collision fraction}
\label{subsec:Calculating-the-merger}
\begin{figure}
\includegraphics[width=1\columnwidth]{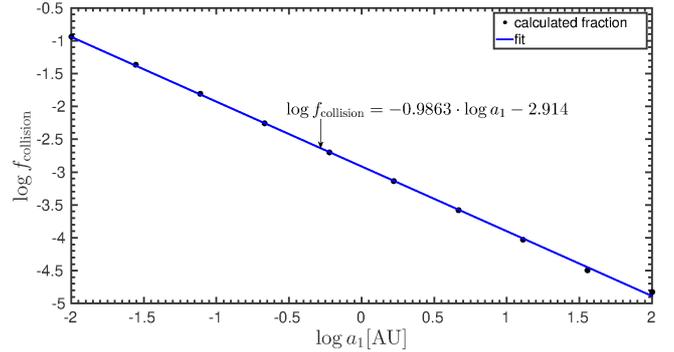}\caption{\label{fig:f_merger} The fraction of systems, $f_{{\rm collide}}$, that the inner binary collide 
during the resonant phase (IMS) as a function of the initial SMA, $a_{1}$.
For every $a_{1}$ we simulated $10^{6}$ binary-single scattering
experiments; for each experiment we use $N_{{\rm IMS}}=20$ scattering
events in which we randomize the temporary binary orbital elements (see
text) and check if this temporary IMS leads to a merger. Black dots,
the calculated fraction for direct collision. Blue solid
line, the fit to a power-law.}
\end{figure}

We execute a numerical calculation to find the fraction of
systems that the inner binary collides during the IMS as a function of ,
$a_{1}$. We sample 10 values of $a_{1}$ equally spaced in log from
$\left(10^{-2}{\rm AU},10^{2}{\rm AU}\right)$. For each value of
$a_{1}$ we randomize $N_{{\rm tot}}=10^{6}$ ``scattering experiments''. 
These experiments are not N-body simulations, but Monte-Carlo approach.
For each scattering experiment we set the number of binary-single
encounters to be $N_{{\rm IMS}}=20$. For each encounter a temporary
binary is created and bound to a third WD on a Keplerian orbit.The temporary binary orbital properties, $a_{{\rm IMS}}$
is drawn uniformly from $\left(a'_{{\rm L}},a'_{{\rm U}}\right)$
see equations (\ref{eq:a_L}) and (\ref{eq:a_U}), and the eccentricity,
$e_{{\rm ecc}}$ is drawn from a thermal distribution. Next we calculate
the temporary pericenter distance $q'=a'\text{\ensuremath{\left(1-e'\right)}}$
and compare it to the combine radii of the two WDs, $2R_{{\rm WD}}$.
If $q'\le2R_{{\rm WD}}$ we count it as a collision and a source of Type
Ia SN. If $q'>2R_{{\rm WD}}$ we calculate $t_{{\rm iso}}$, to keep
track on the time evolution and randomize the binary and single again,
until we reach $N_{{\rm IMS}}$ times. In the case of no collision
during the resonant phase we check the final end state, which have
different orbital parameters distributions, see subsection \ref{subsec:Post-resonance-state}.
$f_{{\rm collision}}\left(a_{1}\right)$ is the number of mergers
divided by $N_{{\rm tot}}$. The results are presented in Figure \ref{fig:f_merger}.
We found a power law relation between $f_{{\rm collision}}$ and $a_{1}$,
the exact fitted function is 
\begin{equation}
f_{{\rm collision}}\left(a_{1}\right)=0.0114\times a_{1}^{-0.954}.\label{eq:merger_fit}
\end{equation}

\subsubsection{Collision time}
\label{subsec:Collision_time}
\begin{figure}

\includegraphics[width=1\columnwidth]{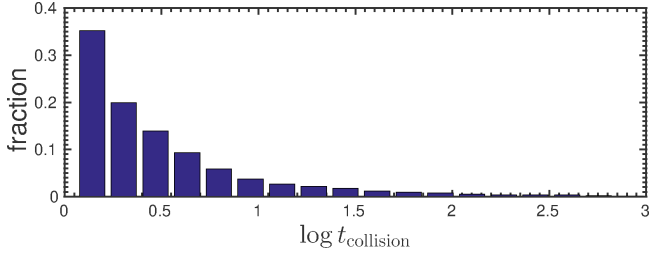}\caption{\label{fig:collision time}The collision time from the onset of the
dynamical instability. The collision times are all much shorter than
the interaction time $t_{{\rm enc}}$ therefore one can treat any
collision as prompt with the turn of the triple to be unstable.}

\end{figure}

Here we show that a collision during the instability phase happens
quickly, on a dynamical timescale, with respect to the evolution time.
Therefore, one can ignore the time elapsed since the beginning of
the instability phase until a direct collision occurs between two
of the WDs. In figure \ref{fig:collision time} we show the distribution
of collision time since the beginning of the resonant stage, it is
evident from the figure that, unsurprisingly, the collision happens
within $\sim1000{\rm yrs}$ since the first binary-single encounter,
hence for the rest of the paper we consider a collision (i.e. SN)
promptly when the system enters the instability stage.

\subsection{Collision in the post-resonance phase}
\label{subsec:Post-resonance-state}
\begin{figure}
\includegraphics[width=1\columnwidth]{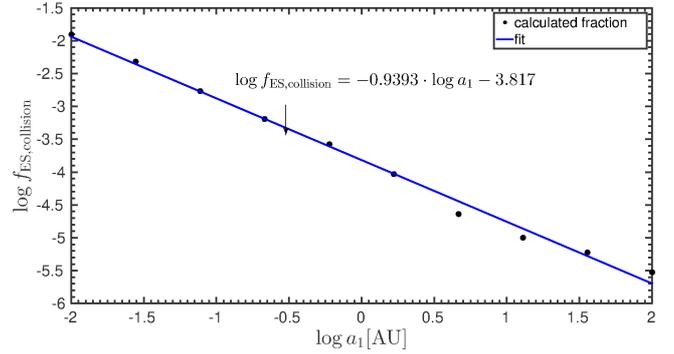}\caption{\label{fig:ES_coll}The fraction of systems that collide at the endstate
of the chaotic dynamics, $f_{{\rm EX,collide}}$ as a function of
the initial SMA, $a_{1}$. For every $a_{1}$ we simulated $10^{6}$
binary-single scattering experiments; Black dots, the calculated fraction for direct collision. Blue solid line,
the fit to a power law.}

\end{figure}

Here we describe the endstate of the resonant interaction where a
tight binary is formed, with a SMA smaller than the initial one,
$a_{{\rm ES}}<a_{1}$ and a single WD escapes the system.
It was previously shown in \citep{Stone2019b,Samsing2014,Heggie1975} that the
binary energy distribution scales like 
\begin{equation}
E_{{\rm ES}}\propto\left|E_{1}\right|^{-4}
\end{equation}
where $E_{{\rm ES}}$ is the energy of the endstate binary and $E_{1}=-Gm_{1}m_{2}/(2a_{1})$
is the energy of the initial binary. Additionally, the endstate eccentricity,
$e_{{\rm ES}}$ is drawn from thermal distribution \citep{Stone2019b}.
Therefore, for every system that did not collide during the IMS an
endstate binary is formed with $a_{{\rm ES}}\left(E_{{\rm ES}}\right)$
and $e_{{\rm ES}}$. This binary either experiences a direct collision
or merges through GW emission. In figure \ref{fig:ES_coll} we present
the fraction of systems that collide in the post-resonance phase as
a function of the initial SMA. The numerical fit presented in the
figure is a broken power law 
\begin{equation}
f_{{\rm ES,coll}}=1.52\cdot 10^{-4}\times a_{1}{}^{-0.9393}\label{eq:ES_coll}
\end{equation}

\section{SN rates}
\label{sec:SNe-rates}
In what follows we calculate the SN rate from wide systems. In subsection
\ref{subsec:SN-rates-IMS} we calculate the Type IA SN rates originating
from TWD systems via direct collisions. In section \ref{subsec:Mergers-in-theES}
we calculate the rates of DD mergers via GW emission for the endstate
of the instability phase.

\subsection{Type Ia SN rates from TWDs}
\label{subsec:SN-rates-IMS}
In section \ref{subsec:Quantitative-description} we describe the
probability of a wide triple becomes unstable due to interaction with
flyby stars. This probability depends on the local stellar environment
through the local stellar density, $n_{*}$ and the local velocity
dispersion which sets the encounter velocity, $v_{{\rm vec}}.$ Therefore
the host galaxy characteristics plays an important role in the SN
rates from this channel.

We model two type of galaxies, spiral and elliptical. For the spiral
galaxy we take the Milky-Way (MW) to represent all star forming galaxies.

Let 
\begin{equation}
dN_{s}(r)=n_{*s}\left(r\right)\cdot2\pi\cdot r\cdot h\cdot dr\label{eq:dN}
\end{equation}
 \citep{Michaely2016} be the total number of stars in a galaxy region $dr$
with scale height $h$. This region of the galaxy is located at distance $r$ from the center.
The galactic stellar density is modeled by the following function
\begin{equation}
n_{{\rm *s}}\left(r\right)=n_{0}e^{-\left(r-r_{\odot}\right)/R_{l}}\label{eq:MW_galaxy}.
\end{equation}
$n_{{\rm *s}}$ is the local stellar density for \textit{spiral}
galaxy and $n_{0}=0.1{\rm pc}^{-3}$ is the stellar density in the solar neighborhood, $R_{l}=2.6{\rm kpc}$ \citep{Juric2008} 
is the galactic length scale. The mass of the flyby is taken to be $0.6M_{\odot}$, 
the average stellar mass in the galaxy. The velocity dispersion
is chosen to be the velocity dispersion of the flat rotation curve of the
galaxy, $\sigma=50{\rm kms^{-1}}$.

\begin{figure}
\includegraphics[width=1\columnwidth]{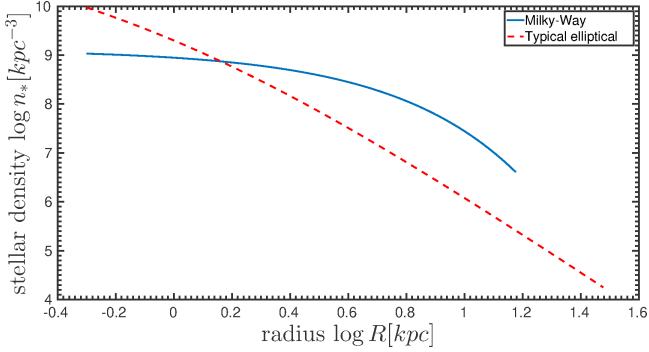}\caption{\label{fig:Numebr-stellar-density}Number stellar density as a function
of distance. The blue solid line represents the Milky-Way galaxy .
The red dashed line represents a typical elliptical galaxy. Figure taken from \citet{Michaely2020}}

\end{figure}

For an elliptical galaxy model we take density profile from \citep{Hernquist1990}
and translate it to stellar density given an average stellar mass
of $0.6M_{\odot}.$
\begin{equation}
n_{{\rm *e}}\left(r\right)=\frac{M_{{\rm galaxy}}}{2\pi r}\frac{r_{*}}{\left(r+r_{*}\right)^{3}}\label{eq:elliptical}
\end{equation}
where $n_{{\rm *e}}$ is the stellar density for \textit{elliptical}
galaxy and $r_{*}=1{\rm kpc}$ is the scale length of the galaxy,
$M_{{\rm galaxy}}=10^{11}M_{\odot}$ is the total stellar (and not
total) mass of the galaxy. Therefore 
\begin{equation}
{\rm dN_{{\rm e}}\left(r\right)}=\frac{n_{{\rm *e}}}{\left\langle m\right\rangle }{\rm dV}
\end{equation}
is the number of stars inside a local volume $dV$ are a distance
$r$ from the center and $\left\langle m\right\rangle $ is the average
stellar mass of the galaxy. The velocity dispersion for a typical
elliptical galaxy is $\sigma=160{\rm kms^{-1}}.$ Figure \ref{fig:Numebr-stellar-density}
shows the stellar density of the two prototypes of galaxies.

Now we estimate the fraction of TWDs out of the stellar population,
$f_{{\rm TWD}}.$ We assume all stars with mass in the range of $1M_{{\rm \odot}}-8M_{\odot}$
become WDs. Given the Kroupa initial mass function \citep{Kroupa2001}
a fraction of $f_{{\rm primary}}\approx0.1$ out of all stars are
in the range of $1M_{{\rm \odot}}-8M_{\odot}$, e.g. will evolve to
become WDs in $10{\rm Gyrs}$ and actually less than that when accounting
for binary stellar evolution. In order to estimate the binary companion
mass we use uniform mass ratio distribution \citep{Moe2016}, $Q_{{\rm inner}}\in\left(0.1,1\right)$.
For the third star we define the outer mass ratio $Q_{{\rm outer}}=m_{3}/\left(m_{1}+m_{2}\right)$
and its value is distributed from a power law distribution \citep{Moe2016}
$f_{Q_{{\rm outer}}}\propto Q_{{\rm outer}}^{-2}$ and $Q_{{\rm outer}}\in\left(0.1,1\right)$,
which is similar to random pairing to the initial mass function. Given
these distributions we get that the fraction of secondaries in the
range of producing WDs is $f_{{\rm secondary}}\approx0.45$, for the
tertiaries is $f_{{\rm tertiary}}\approx0.42$. The fraction of triples
is chosen to be $f_{{\rm triple}}=0.2$ and the fraction of wide outer
binaries greater than $1000{\rm AU}$ from a log-uniform distribution,
$f_{a_{2}}$, is $f_{{\rm wide}}=0.2.$ Combining these estimations
we get 
\begin{equation}
f_{{\rm TWD}}=f_{{\rm primary}}\times f_{{\rm secondary}}\times f_{{\rm tertiary}}\times f_{{\rm triple}}\approx3.8\times10^{-3}
\end{equation}
 out of which only $f_{{\rm wide}}$ are in wide configuration, hence
\begin{equation}
f_{{\rm model}}=f_{{\rm TWD}}\times f_{{\rm wide}}=7.6\times10^{-4}
\end{equation}
is the fraction of the stellar population in a wide TWD configuration.

We note here, and expand in the discussion, that this is a simplification
of a very complex estimation. In order to correctly calculate the
wide TWD systems out of a certain stellar population one need to consider
both single and binary stellar evolution. This would modify the initial
SMA and eccentricity distributions whilst change the masses. Specifically,
common envelope evolution \citep{Ivanova2013} modifies the inner
SMA and even the outer SMA due to mass loss from the inner binary
\citep{Michaely2019c,Igoshev2020}.

Next we calculate the total SN rate for the MW-like galaxy and an
typical elliptical galaxy. The rate, $\Gamma$ is given by integrating
the loss cone (\ref{eq:empty}) and (\ref{eq:full}) for all outer
SMAs, $a_{2}$ between $10^{3}-10^{5}{\rm AU}$, the local stellar
density in the galaxy $n_{*}$ from equations (\ref{eq:MW_galaxy})
and (\ref{eq:elliptical}). In order to integrate the inner binary
SMA we use the following limits $10^{-2}\left(10^{-1}\right)-10^{2}{\rm AU}$:
\begin{equation}
\Gamma=\int\int\int\frac{L_{{\rm collision}}\left(a_{1},a_{2},n_{*}\right)}{10{\rm Gyr}}da_{1}da_{2}dN\left(r\right)\label{eq:the_integral}
\end{equation}
where $L_{{\rm collision}}\equiv L\left(a_{1},a_{2},n_{*}\right)f_{a_{1}}f_{a_{2}}f_{{\rm model}}f_{{\rm collision}}$
and we define 
\begin{equation}
dL\equiv\frac{L_{{\rm collision}}\left(a_{1},a_{2},n_{*}\right)}{10{\rm Gyr}}da_{1}da_{2}dN\left(r\right),
\end{equation}
and write the integral for the MW-like galaxy 
\begin{equation}
\Gamma_{{\rm MW}}=\int_{{\rm 0.5kpc}}^{{\rm 15kpc}}\int_{{\rm 10^{3}AU}}^{10^{5}{\rm AU}}\int_{{\rm 10^{-2}\left(10^{-1}\right)AU}}^{10^{2}{\rm AU}}{\rm dL}\approx2\ \left(0.1\right)\times10^{-5}{\rm yr^{-1}}
\end{equation}
and for a typical elliptical 
\begin{equation}
\Gamma_{{\rm elliptical}}=\int_{{\rm 0.1kpc}}^{{\rm 30kpc}}\int_{{\rm 10^{3}AU}}^{10^{5}{\rm AU}}\int_{{\rm 10^{-2}\left(10^{-1}\right)AU}}^{10^{2}{\rm AU}}{\rm dL}\approx3.4\ \left(0.2\right)\times10^{-5}{\rm yr^{-1}.}
\end{equation}
These results are averaged on a $10{\rm Gyr}$ lifetime of the galaxies,
in section \ref{subsec:Delay-time-distribution} we discuss the delay-time
distribution of these collision.

In the case where a direct collision did not occur during the IMS
phase the triple is disrupted and a compact binary is formed. We calculate
the collision rate by using, $f_{{\rm ES,col}}$ from (\ref{eq:ES_coll})
instead of $f_{{\rm collision}}$ to get a SN rate for the MW-like
galaxy 
\begin{equation}
\Gamma_{{\rm ES,MW}}=\int_{{\rm 0.5kpc}}^{{\rm 15kpc}}\int_{{\rm 10^{3}AU}}^{10^{5}{\rm AU}}\int_{{\rm 10^{-2}\left(10^{-1}\right)AU}}^{10^{2}{\rm AU}}{\rm dL}\approx2\ \left(0.12\right)\times10^{-6}{\rm yr^{-1}}
\end{equation}
 and for the elliptical galaxy a rate of 
\begin{equation}
\Gamma_{{\rm ES,elliptical}}=\int_{{\rm 0.1kpc}}^{{\rm 30kpc}}\int_{{\rm 10^{3}AU}}^{10^{5}{\rm AU}}\int_{{\rm 10^{-2}\left(10^{-1}\right)AU}}^{10^{2}{\rm AU}}{\rm dL}\approx3.4\ \left(0.21\right)\times10^{-6}{\rm yr^{-1}.}
\end{equation}
 The total rate for the case of $a_{1}\in\left(10^{-2}\left(10^{-1}\right){\rm AU},10^{2}{\rm AU}\right)$,
over a galactic lifetime is for MW-like galaxy is $\sim2.2\ \left(0.11\right)\times10^{-5}{\rm yr^{-1}}$
and for elliptical galaxies $\sim3.8\ \left(0.22\right)\times10^{-5}{\rm yr^{-1}}.$
Comparing that to the observed rate over the lifetime of a typical
galaxy $10^{-3}{\rm yr^{-1}}$ this channel may explain $\sim0.1-2\%$
of all type Ia SNe in MW-like galaxies and $\sim0.2-4\%$ in elliptical
galaxies.

We note here and expand in the discussion that the uncertainty in
expected collision rate, hence SN rate, is due to the uncertain structure
of the triple systems, specifically the inner binary SMA distribution.
The common envelope evolution may change the distribution of the SMA
or even cause the inner binary itself to merge.

\subsection{Mergers in the post-resonance phase}
\label{subsec:Mergers-in-theES}
In the case where no direct collision occurred in the instability
stage and imminently after the break-up of the triple into a compact
WD binary and an escaper WD. We are left with two WD in a relativity
close binary and eccentricity. These binaries omit GW and spiral in
to eventually merge similarly to the classical DD scenario. Here we
calculate the rate of these occurrences.

The merger timescale via GW emission for eccentric binaries is given
from \citep{Pet64} 
\begin{equation}
t_{{\rm merger}}\approx\frac{768}{425}T_{c}\text{\ensuremath{\left(a\right)}}\left(1-e^{2}\right)^{7/2}\label{eq:t_merger}
\end{equation}
where $T_{c}=a^{4}/\beta$ is the merger timescale for a circular
orbit and $\beta=64G^{3}m_{i}m_{j}\left(m_{i}+m_{j}\right)/\left(5c^{2}\right)$.
Here $m_{i/j}$ are the indexes of the random two WDs that ended up
as the surviving compact binary and $c$ is the speed of light. If
$t_{{\rm merger}}<10^{10}{\rm yr}$ we flag this systems as a DD inspiral.

\begin{figure}
\includegraphics[width=1\columnwidth]{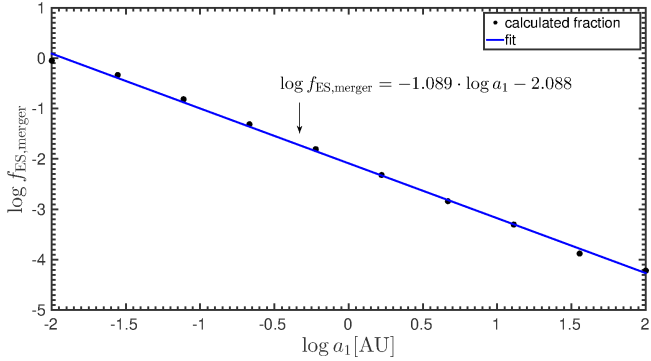}\caption{\label{fig:f_ES_merger}The fraction of systems that merge within
$10^{10}{\rm yr}$ at the endstate of the chaotic dynamics, $f_{{\rm ES,merer}}$
as a function of the initial SMA, $a_{1}$. For every $a_{1}$ we
simulated $10^{5}$ binary-single scattering experiments; Black dots,
the calculated fraction of endstate mergers from our numerical experiment.
Blue solid line, the best fit to a power law.}
\end{figure}

In figure \ref{fig:f_ES_merger} we present the calculated fraction
of DD merger in the post-resonance phase. We found a broken power-law
fit to the merger fraction 
\begin{equation}
f_{{\rm ES,merger}}=0.008\times a_{1}^{-1.089}.
\end{equation}

Equipped with the functional form of the merger fraction we plug these
in equation (\ref{eq:the_integral}) to get the following rate for
both types of galaxies and initial conditions:

\begin{equation}
\Gamma_{{\rm ES,MW}}=\int_{{\rm 0.5kpc}}^{{\rm 15kpc}}\int_{{\rm 10^{3}AU}}^{10^{5}{\rm AU}}\int_{{\rm 10^{-2}\left(10^{-1}\right)AU}}^{10^{2}{\rm AU}}{\rm dL}\approx2.2\ \left(0.1\right)\times10^{-4}{\rm yr^{-1}}
\end{equation}
 and for the elliptical galaxy 
\begin{equation}
\Gamma_{{\rm ES,elliptical}}=\int_{{\rm 0.1kpc}}^{{\rm 30kpc}}\int_{{\rm 10^{3}AU}}^{10^{5}{\rm AU}}\int_{{\rm 10^{-2}\left(10^{-1}\right)AU}}^{10^{2}{\rm AU}}{\rm dL}\approx3.6\ \left(0.16\right)\times10^{-4}{\rm yr^{-1}.}
\end{equation}
Which account for $\sim1-36\%$ of the Type Ia rates. This is an addition
to the classical DD rate because these systems originate from binaries
that their GW inspiral time is greater than Hubble time.

\section{Discussion}
\label{sec:Discussion}
\subsection{Model assumptions}

The mathematical model and calculation we present here is based on
several assumptions, in the following we address them.

\textbf{WD masses}. In this manuscript we use WD mass of $0.6M_{\odot}$
this is a simplification of single stellar evolution. Different WD
masses correspond to different WD radii via the mass radius relation
(\ref{eq:M-R_relation}) and therefore different cross-section for
a direct collision. Additionally, if the triple consists of significantly
different masses the assumption of $N_{{\rm IMS}}=20$ breaks down
and a different treatment should be done.
Moreover, different WD masses correspond to different triple evolution timescale. 
The importance of this issue is with regarding of the interaction of the triples with flybys during the MS lifetime, and 
with the time where actually the TWD is formed. A dedicated population synthesis will shed light on these issues.   

\textbf{Triple fraction and orbital elements}. A key ingredient in
this model in order to calculated the SN rate is the fraction of triple
systems out of the stellar population, $f_{{\rm model}}$ and the
distributions of the SMAs and eccentricities. The fractions we use
in order to estimate $f_{{\rm model}}$ are taken from \citet{Moe2016}
which describe MS binary stars and not WD binaries. The same holds
for the SMA and eccentricity distributions which are motivated from
the MS systems.

Single and binary stellar evolution effect both the SMA and eccentricity
for each system. Mass-loss due to stellar evolution (slow mass-loss)
leads to the expansion of the system's SMA while keeping the eccentricity
constant. However, complex binary interaction, e.g. tidal interaction,
mass transfer, CEE, might change both the SMAs and the inner and outer
eccentricity considerably and might even disrupt the binaries \citep{Michaely2019,Michaely2020}.
As we present in section \ref{subsec:SN-rates-IMS} the uncertainty
in the SN rate is primarily effected by the inner SMA distribution
and the lower boundary of the inner SMA, which is a result of CEE.
In a future study we intend to account for the dynamics described
here with a population synthesis study that account for these interaction
in order to get a more accurate description of the SN rates.

\subsubsection{Two WDs in a triple system}

In this study we focused on TWD. A complementary fraction of the population
is a triple system with two WDs and a low mass stellar companion,
$<1M_{\odot}$. In these systems the dynamics are similar with the
scenario presents here because the third star have to be a low mass
stellar companion hence similar in mass with the other two WDs. The
only difference is that the low mass stellar companion radius is orders
of magnitude greater than the WD radius. Therefore, the most probable
outcome of the resonant phase is a direct collision between a WD and
the low mass star. This collision occurs in the presence of a bound
WD in a wide and eccentric orbit. This interesting scenario is not
studied here and will be studies elsewhere.

\subsubsection{Delay time distribution (DTD)}
\label{subsec:Delay-time-distribution}
An important observable of Type Ia SNe is the DTD, $d{\scriptscriptstyle N}/dt$.
The DTD is the hypothetical rate of Type Ia SNe that follows a quick
star formation. It have been well established that the observed DTD
is proportional to $1/t$ where $t$ is the time since star formation
\citep{Maoz2014}. In this section we try to estimate the DTD profile
of the dynamical scenario described here.

\begin{figure}
\includegraphics[width=1\columnwidth]{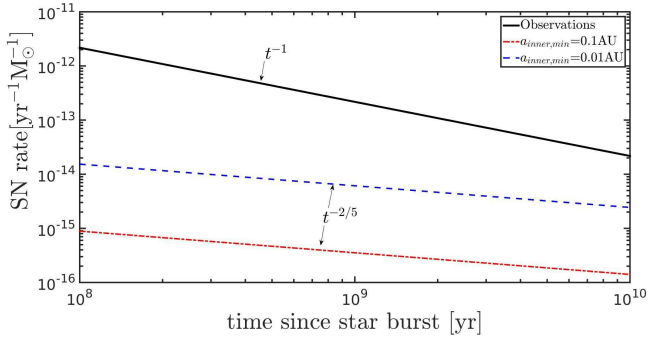}\caption{\label{fig:The-delay-time-distribution.}The delay-time distribution.
Black solid line is the observed DTD, corresponds to a $10^{-3}{\rm yr^{-1}}$
SN rate for a $10^{10}M_{\odot}$ galaxy. Blue dashed line is the
estimated DTD for the 2WDC channel described here for the upper rate
of $3.8\times10^{-5}{\rm yr^{-1}}$ while the Red dashed-dot line
is the DTD of the lower rate of $2.2\times10^{-6}{\rm yr^{-1}}.$ These
DTD are calculated while ignoring the wide binary ionization process
(described in the text) therefore should be treated as a upper limit.}

\end{figure}

The DTD is governed by the numbers of available systems to collide
as a function of time. This in turn is a function of the initial mass
function and the stellar evolution time for each mass. Additionally,
as stated in previous work, the rate of flyby interaction is constant
in time if one disregards binary ionization. Therefore we can write
the following dependency
\begin{equation}
\frac{d{\scriptscriptstyle N}}{dt}\propto\frac{d{\scriptscriptstyle N}}{dm}\frac{dm}{dt}.
\end{equation}
In order to estimate the first term, we examine the star with lowest mass in the initial triple. This star 
will be the last to turn into a WD, hence determines the time since start formation where the TWD is formed. Given 
the population we simulated in \ref{subsec:SN-rates-IMS} in order to calculate the triple fraction out of a given stellar population, 
one can find that the lowest stellar mass in a TWD progenitor scale with stellar mass, $d{\scriptscriptstyle N}/dm=m^{-2.75}$. This is steeper than 
for both Kroupa and Salpeter \citep{Kroupa2001,Salpeter1955} IMFs.
The second term is just the MS life time, $t_{{\rm MS}}$, i.e. the
time that takes a MS star evolve into a WD, $dm/dt=t^{-4/3}$. For
these simplifying assumption we get 
\begin{equation}
\frac{d{\scriptscriptstyle N}}{dt}\propto t^{2.75/3}t^{-4/3}=t^{-0.41}\approx t^{-2/5}.
\end{equation}
As mentioned above the observed DTD of Type SNe is $t^{-1},$ this
implies the as time progress the relative importance of the 2WDC channel
increases and for late times this channel dominate over other channel
see figure \ref{fig:The-delay-time-distribution.}. After $10^{10}\rm yr$ roughly up to $10\%$ of all Type Ia SNe originate from the collision channel. We emphasize that
in this calculation we neglected the ionization of wide binaries due
to flyby interaction in the field. Equation (\ref{eq:ionization})
shows the complex dependencies of the ionization of a wide binary,
specifically the SMA, $a$ and the local stellar density. Therefore,
one should treat figure \ref{fig:The-delay-time-distribution.} as
an upper limit due to ionizations in later time as seem in equations
(\ref{eq:empty}) and (\ref{eq:full}).

\subsection{Collisions and mergers from wide WD binaries}

One can imagine a scenario where two WDs collide from a wide binary
configuration. This scenario is far less efficient because the loss
cone in the binary case, $F_{{\rm q,2}}$ is orders of magnitude less
than the loss cone in the triple case, $F_{{\rm q,3}}$ (\ref{eq:Fq}):
\begin{equation}
F_{{\rm q,2}}=\frac{2R_{{\rm WD}}}{a_{2}}\sim\frac{4\times10^{-5}{\rm AU}}{10^{4}{\rm AU}}\ll\frac{2\times10^{-1}{\rm AU}}{10^{4}{\rm AU}}=\frac{2a_{1}}{a_{2}}=F_{{\rm q,3}}.
\end{equation}
However, binaries one order of magnitude more frequent than triples
therefore we find interest in calculating their rate in the future.

Moreover, similar to the scenario presented in \citep{Michaely2019},
the binary eccentricity can be excited to sufficiently high values
so that the GW merger time, $t_{{\rm merger}}\left(a,e\right)$, is
shorter than the time between stellar encounters, $t_{{\rm enc}}$.
The time between encounter timescale is given by \citep{Michaely2019}
\begin{equation}
t_{{\rm enc}}=\frac{1}{f}=\left(n_{*}\sigma v_{{\rm enc}}\right).^{-1}
\end{equation}
For this channel, the loss cone is 
\begin{equation}
F_{{\rm q,GW}}=\left(\frac{\beta t_{{\rm enc}}}{a_{2}^{4}}\right)^{-1}.
\end{equation}
We will dedicate a future research for wide binary WDs are source
for Type Ia SNe for completeness, and wide binaries consist of WD
and a stellar companion as a source of cataclysmic variables.

\section{Summary}
\label{sec:Summary}
In this manuscript we explore the dynamical channel in which a TWD
becomes dynamically unstable due to interactions with field stars.
As a results the previously stable triple acts chaotically and experiences
multiple binary-single encounters. In every such encounter there is
a chance that the temporary inner binary would collide and result
as a Type Ia SN. In the case where the triple survives the multiple
binary-single encounters, the systems breaks down to a compact binary
and an escaper WD. The compact binary either collide on its first
orbit or merge via GW emission in much later time. If the inspiral
time is shorter than the Hubble time, this system will merge, similar
to the DD scenario.

We find that this dynamical channel that leads to a two WDs collision,
may explain $2-36\%$ of the observed Type Ia rates from the merger of two WDs and $.1-4\%$ from the direct collision channel, with the following caveats: 
the uncertainty of the inner binary evolution and the ionization of the outer binary. In this study 
we neglected the issue of the morphology of the remnant which is observed to be rather spherical in close by SN remnants and not yet well predicted in 
the 2WDC models.

\section*{Acknowledgments}

EM thanks Todd Thompson for enlightening comments, Hagai Perets and Noam Soker for interesting discussions together with the CTC (center of theory and computation) of the University of Maryland for financial support. 

\textbf{Data availability} The data underlying this article will be shared on reasonable request to the corresponding author.
\bibliographystyle{mnras}
\bibliography{TBH}

\end{document}